# SPECTROSCOPY OF SUPERDEFORMED Pb ISOTOPES*

H. Hübel

Institut für Strahlen- und Kernphysik, University of Bonn, Nussallee 14-16, D-53115 Bonn, Germany

Using the spectrometer arrays GAMMASPHERE and EUROBALL a detailed spectroscopy of superdeformed $^{193}$Pb and $^{196}$Pb has been performed. In $^{193}$Pb nine superdeformed bands are now known and their quasiparticle assignments are discussed. In $^{196}$Pb evidence for an octupole vibrational excitation is presented.

## 1. Introduction

In the mass region around A = 190 about 70 superdeformed (SD) rotational bands are known in 25 different nuclei [1]. In many cases several bands have been found which are built on excited states in the SD potential energy minimum. The nature of these excitations has been the subject of experimental and theoretical investigations in recent years. In the odd-even and odd-odd nuclei it seems clear that the low-lying states are quasiparticle excitations. However, in the even-even isotopes the nature of the excited SD bands in this mass region is less clear. In almost all cases configuration assignments are difficult since the bands are not linked to lower-lying known states.

With the large highly-efficient $\gamma$-ray coincidence spectrometers GAMMASPHERE [2] and EUROBALL [3], it is possible to perform a detailed spectroscopy of SD nuclei and to collect sufficient experimental evidence to make reliable configuration assignments. In this contribution, two such investigations of Pb isotopes are selected as examples, the odd-A nucleus $^{193}$Pb [4] and the even-even nucleus $^{196}$Pb [5]. In $^{193}$Pb a total of nine

---

SD bands are now known and quasiparticle assignments are made to all of them. In $^{196}$Pb, where an excited band was known to decay to the yrast SD band [6], evidence is presented that the connecting transitions are strongly enhanced electric dipoles. This suggests that the excited band is built on an octupole vibration.

## 2. Quasiparticle Excitations in SD $^{193}$Pb

In the previous work on SD $^{193}$Pb six bands, the yrast pair plus two sets of excited signature-partner bands, were identified [7, 8]. In a new investigation, our group has studied high-spin states in this nucleus [4] in an experiment performed at LBNL using the Gammasphere array. A beam of $^{30}$Si, provided by the 88-Inch Cyclotron, with an energy of 160 MeV was incident upon a thin, self-supporting foil of $^{168}$Er. Data were collected to tape when a minimum of 4 Ge detectors showed a prompt coincidence, with the secondary condition that 3 of these signals remain after pile-up rejection. After selecting only those events which occurred during the prompt beam burst, a total of $1.7 \times 10^9$ events remained, which unpacked to $0.94 \times 10^9$ four– and higher-fold coincidences.

The data were thoroughly searched for regular structures with energy spacing around 35–45 keV and three new bands were found. On the basis of coincidences with the lower-lying ND states, these structures are assigned to $^{193}$Pb. They are referred to as bands 7, 8 and 9 hereafter. Bands 2 to 9 have intensities of 0.38, 0.46, 0.23, 0.15, 0.20, 0.17, 0.14 and 0.05, normalised to an intensity of 1.0 for the yrast SD band 1. Each of these values has an uncertainty of at least 20% due to contributions from the background subtraction [4]. The transition energies of all nine bands are given in Table 1.

None of the nine SD bands in $^{193}$Pb are connected to low-lying known states and, consequently, their excitation energies and spins are not known. As the intensities with which bands 1 through 9 are populated in the nuclear reaction decrease, one can assume that their excitation energies increase. The spins of the bands may be obtained by spin-fitting methods. In the case of the yrast pair (bands 1 and 2), such methods are inappropriate due to their probable intruder nature. Thus, for these two bands we adopt the spins suggested by Hughes *et al.* [7]. For band 9 the results of the spin-fitting methods are ambigous because of the irregularity of the dynamic moment of inertia at low rotational frequency. The spins of the band heads are included in Table 1.

Arguments about the configuration assignments to the bands can primarily be made by comparing experimental and theoretical Routhians, calculated within the framework of the cranked shell model (CSM). As the experimental excitation energies of the bands are not known, the relative position of the Routhians displayed in Fig. 1 are arbitrary; they are ordered according to the intensities of the bands. For the signature-partner bands it was assumed that the energy splitting at $\hbar\omega = 0$ vanishes. The lowest-energy bands are built on the $7_2$ quasineutron intruder orbital (originating from the [761]3/2 Nilsson state). The large signature splitting of the strongly fed bands 1 and 2 supports this assumption. With the spins adopted for bands 1 and 2 (see Table 1) these bands have signature $\alpha = -1/2$ and $+1/2$, respectively. This result is in agreement with the calculation which gives $\alpha = -1/2$ for the favoured signature. For bands 3 and 4 Hughes et al. [7] suggest the [512]5/2 configuration. Our WS calculations predict a higher energy for this orbital. The calculations show that, after the yrast pair, the next available quasineutron excitations which exhibit some signature splitting are the bands built on the levels of [642]3/2 origin. Therefore, we assign this configuration to bands 3 and 4. This agrees well with the HF+BCS calculations presented by Ducroux et al. [8]. On the basis of the fitted spins (see Table 1), band 4 (which is favoured in energy at high rotational frequency) has $\alpha = -1/2$.

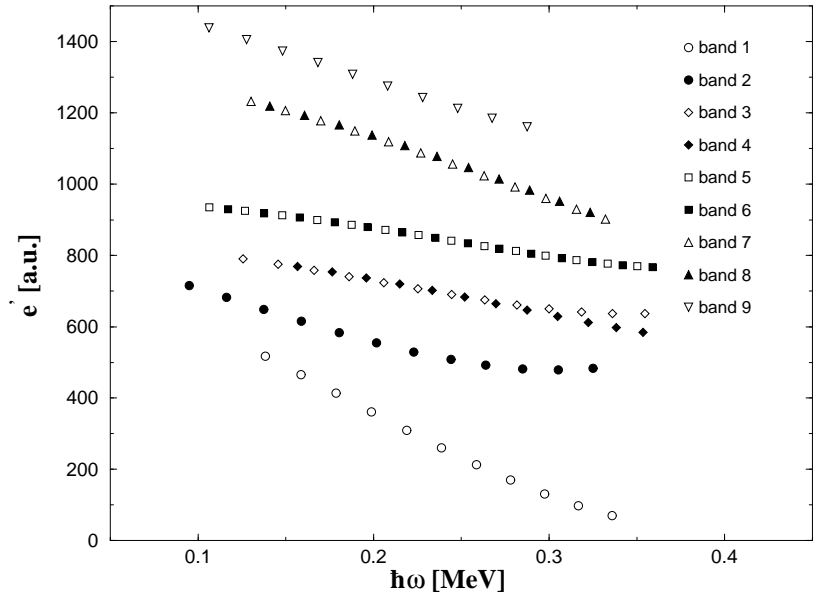

Fig. 1. Experimental Routhians with arbitrary excitation energy. For the signature-partner bands it was assumed that the splitting vanishes at $\hbar\omega = 0$.

For bands 5 and 6 the results of a $g$-factor analysis suggest the [624]9/2 configuration. This orbital appears in our calculations at rather high ex-citation energy at low frequency, but it decreases rapidly in energy with increasing frequency. It is, therefore, populated with reasonable intensity at high spin in the heavy-ion reaction. It should result in a strongly coupled pair of signature partners, showing very little signature splitting over the entire frequency range and exhibiting cross-talk between the bands, in agreement with the experimental observation.

Table 1. Transition energies for all nine SD bands in $^{193}$Pb. The transitions in brackets have been suggested earlier [8], but could not be confirmed in the present data.

| Band 1 (27/2) | Band 2 (17/2) | Band 3 (21/2) | Band 4 (23/2) | Band 5 (17/2) | Band 6 (19/2) | Band 7 (23/2) | Band 8 (25/2) | Band 9 (17/2,19/2) |
|---|---|---|---|---|---|---|---|---|
| 277.0(3) | 190.2(5) | 251.5(6) | [273.0] | 213.2(4) | 234.6(5) | 260.6(7) | 281.8(6) | 212.9(5) |
| 317.3(3) | 232.6(3) | 291.5(3) | 313.4(6) | 253.6(7) | 275.5(5) | 299.8(6) | 321.5(6) | 255.8(5) |
| 357.3(3) | 275.2(3) | 332.4(3) | 353.1(4) | 296.2(5) | 316.2(5) | 340.1(4) | 360.9(5) | 297.3(6) |
| 397.5(3) | 317.9(3) | 372.1(3) | 391.9(3) | 336.1(4) | 355.9(5) | 378.9(5) | 398.5(5) | 336.6(6) |
| 437.8(3) | 360.9(3) | 411.9(3) | 430.0(3) | 375.1(5) | 393.4(5) | 417.2(5) | 435.9(4) | 375.8(5) |
| 477.4(3) | 403.5(3) | 450.6(3) | 467.1(4) | 413.5(5) | 432.8(4) | 454.0(5) | 472.3(6) | 415.9(4) |
| 517.3(4) | 445.9(3) | 488.9(3) | 503.9(4) | 451.2(5) | 470.6(4) | 490.1(5) | 508.1(6) | 455.5(4) |
| 556.1(3) | 488.2(4) | 526.6(4) | 539.5(4) | 488.6(5) | 507.4(4) | 526.1(5) | 543.2(6) | 495.6(6) |
| 594.8(4) | 528.0(5) | 563.4(4) | 575.1(3) | 526.5(5) | 543.5(5) | 561.3(5) | 578.0(5) | 535.4(7) |
| 633.4(5) | 569.8(6) | 599.9(5) | 610.0(5) | 562.2(6) | 579.7(5) | 596.4(6) | 612.0(6) | 575.3(8) |
| 671.8(6) | 610.5(7) | 637.0(5) | 644.5(6) | 596.2(7) | 614.6(7) | 631.1(7) | 646.8(7) | |
| [708.2] | 650.0(7) | 672.2(6) | 676.4(6) | 631.3(8) | 649.5(5) | 664.2(7) | | |
| | [689.8] | 709.2(7) | 707.2(8) | 666.8(9) | 684.0(6) | | | |
| | | | | 700.5(8) | 717.9(7) | | | |

In addition to the quasiparticle Routhians assigned to bands 1 to 6, the calculations show several further Routhians close to the Fermi surface which could be populated with similar intensities to bands 5 and 6. These are the Routhians based on the [640]1/2, [505]11/2, [521]1/2, [512]5/2 and [752]5/2 Nilsson orbitals [4]. They display either a large signature splitting down to the lowest frequencies ([640]1/2, [521]1/2 and [752]5/2), a very small signature splitting at high frequency ([512]5/2) or no splitting at all ([505]11/2). For the new bands 7 and 8 we find a small signature splitting at higher frequencies. Thus, the logical assumption seems to be that the two new bands 7 and 8 are based on the [512]5/2 quasineutron excitation. The new band 9, for which we do not observe a signature partner, may be associated with the favoured signature of any of the calculated Routhians with a large splitting. However, the dynamic moment of inertia $\Im^{(2)}$ of band 9 does not show an increase with increasing rotational frequency similar to bands 3 to 8. This is a behaviour which is typical for intruder orbitals in this mass region. A look at the relative intensity of band 9, which - with about 5% of the yrast SD band is even smaller than the relative intensity of band 8 - suggests that this band lies higher in excitation energy than the other eight bands in $^{193}$Pb. Therefore, the $7_3$ quasineutron intruder orbital originating from the [752]5/2 Nilsson orbital is a plausible candidate for the configuration of band 9.

## 3. Octupole Vibration in SD $^{196}$Pb

The new experiment to study SD $^{196}$Pb [5] was performed at Strasbourg using the EUROBALL IV $\gamma$-ray detector array. High-spin states were populated in the $^{170}$Er($^{30}$Si,4n)$^{196}$Pb reaction at a beam energy of 144 MeV. The beam, provided by the Vivitron Tandem accelerator, was incident upon a Au-backed $^{170}$Er target of 1.65 mg/cm$^2$ thickness.

In a previous study of SD $^{196}$Pb [6] $\gamma$ rays were observed to connect states of the second excited band (band 3) to states of the yrast band (band 1), see Fig. 2. A directional correlation (DCO) analysis indicated that these transitions are dipoles. In our experiment the linear polarization of the linking transitions was measured using the Clover detectors of the EUROBALL array as Compton polarimeters [5]. The high efficiency of the array was necessary to measure the linear polarization of the weak linking transitions. Their intensity is of the order of $10^{-4}$ of the intensity of the 4n reaction channel.

Two linear polarization spectra are shown in Fig. 3 for horizontally and vertically scattered $\gamma$ rays, respectively. In-band transitions are marked by an asterisk, transitions which belong to the normal-deformed (ND) level scheme of $^{196}$Pb are labelled y. The high-energy region of each spectrum is expanded in the insets. The energies of the inter-band transitions lie between 935 and 946 keV (see Fig. 2). The three transitions around 945 keV cannot be resolved in the $\gamma$-ray spectra and appear as one broadened peak marked by the arrow in Fig. 3. As the multipolarities of these transitions are necessarily the same, the summed peak has been used to make a combined measurement of the polarization. This increases the statistical accuracy of the measurement.

Using the experimentally derived sensitivity [5], the polarization $P$ has been calculated for the group of inter-band transitions (the sum peak at 945 keV), for in-band E2 transitions between the SD states and for several ND transitions of known multipolarity. The results are shown in Fig. 4. Although the uncertainty in the value for the linking transitions, $P(945 \text{ keV}) = 1.14 \pm 0.51$, is quite large due to their low intensity, we can strongly assert that they are electric transitions. This result for $^{196}$Pb represents only the second case in the A = 190 region (after $^{190}$Hg [9]) in which the E1 character of the transitions connecting an excited SD band to the yrast SD band has been determined experimentally.

Since no lifetime measurement is available from which the E1 transition probability could be obtained, an estimate of the B(E1) values may be obtained from the inter-band (E1) to the in-band (E2) branching ratios. In the analysis the same quadrupole moment as has been deduced for the yrast SD band ($Q_t = (19.5\pm0.4)$ $b$) from a lifetime measurement [10] has been used for band 3. This gives B(E1) = $(1.0 \pm 0.2) \cdot 10^{-4}$ Wu. This value is much larger than E1-transition strengths normally found between excited quasiparticle states.

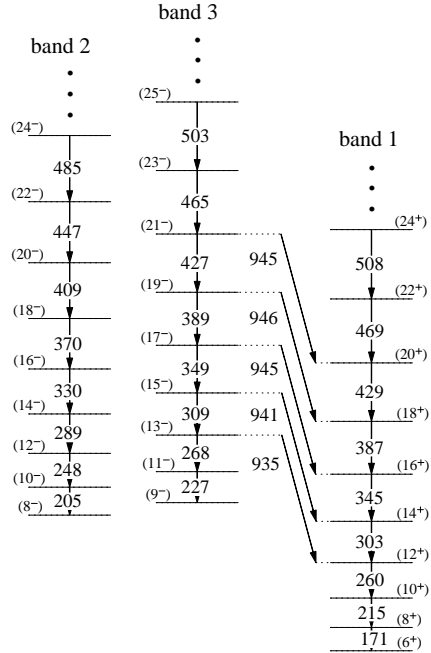

Fig. 2. Partial level scheme of $^{196}$Pb showing the decay of the second excited SD band (band 3) to the yrast SD band (band 1). Bands 2 and 3 are thought to be signature partners with signature $\alpha = 0$ and 1, respectively [6]. The transition energies are given in keV.

The most natural explanation of the enhanced E1 transitions is the presence of octupole correlations. Recent RPA calculations by Nakatsukasa et al. [11] suggest that the lowest excitations in $^{196}$Pb are the K = 2 octupole states. The signature partners based on this state are predicted to lie about 850 keV above the SD vacuum at spin $I = 0$. This is remarkably close to the extrapolated excitation energy of band 3 relative to band 1. The calculations also predict that only the $\alpha = 1$ signature of this mode will exhibit a strong decay branch to the yrast SD band, as is observed experimentally for band 3. Such decay becomes possible only through mixing with lower-K wave functions. The degree of mixing for the K = 2 signature $\alpha = 1$ is predicted to be much greater than for the $\alpha = 0$ state. Band 2, whose energies lie at the midpoints of band 3, thus seems a likely candidate for the K = 2, $\alpha = 0$ octupole state.

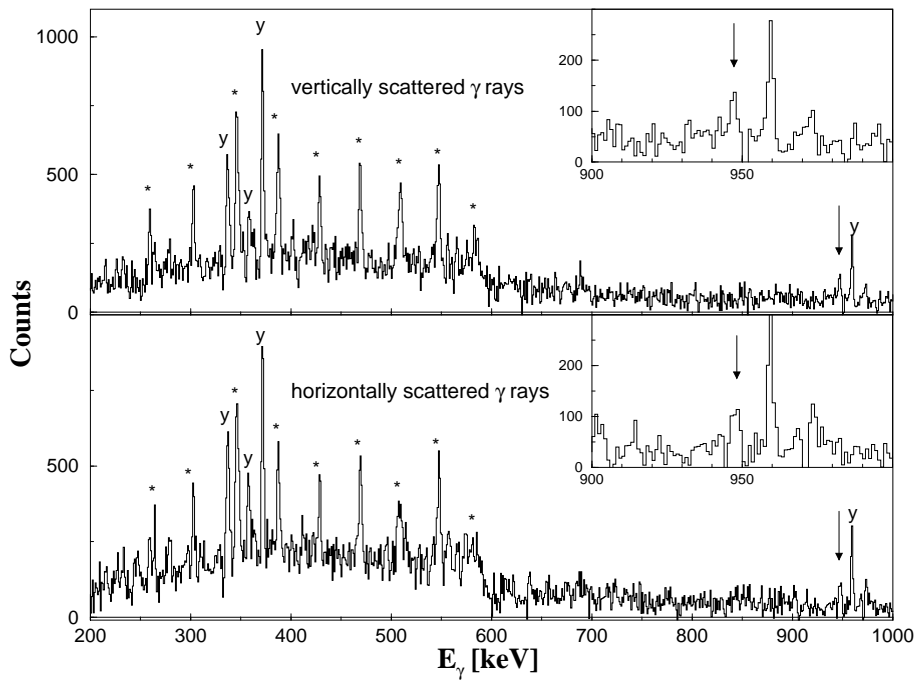

Fig. 3. Projection spectra created from matrices with vertically (upper part) and horizontally (lower part) scattered $\gamma$ rays. See text for details. The group of transitions around 945 keV connecting band 3 to band 1 is marked by the arrows.

The situation is similar to $^{190}$Hg, where SD band 2 is also explained as the K = 2, $\alpha = 1$ octupole vibration [9, 11, 12]. Experimental and calculated B(E1) values for these states are compared in Table 2. The same quantities are included for $^{194}$Hg, the only other case in which transitions have been observed between excited and yrast SD bands in this mass region. Although there is no direct experimental evidence that these transitions are electric dipoles, it is reasonable to assume their E1 multipolarity from the observed decay of the excited band [13]. In all three cases, the E1 transition probabilties are unusually large.

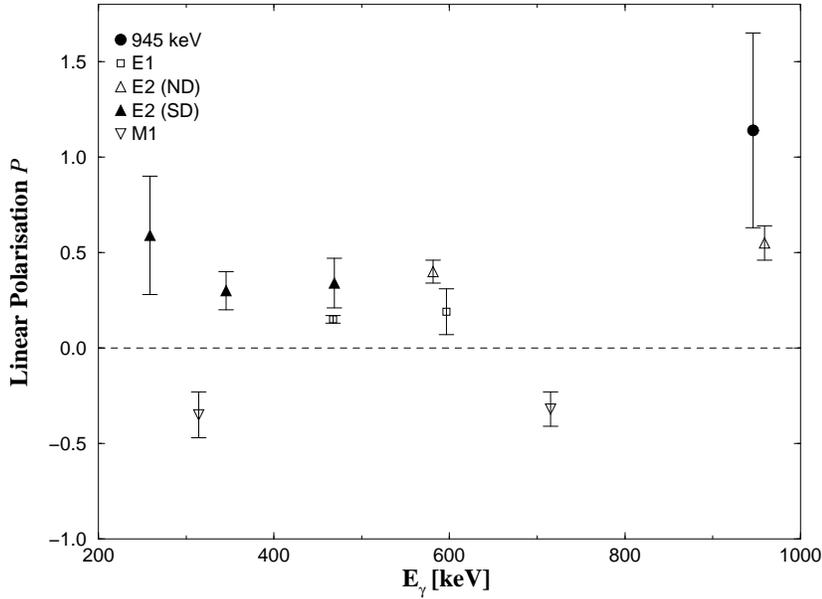

Fig. 4. Polarization of the SD inter-band transitions (full circle), SD in-band transitions (full triangles) and transitions from the ND decay scheme of known multipolarity (open symbols). For pure stretched electric transitions $P$ is positive, for magnetic transitions it is negative.

Table 2. Experimental and calculated B(E1) values (in Wu) for transitions between excited ($K = 2$, $\alpha = 1$) and yrast ($K = 0$, $\alpha = 0$) SD bands in the A = 190 region.

|  | $^{196}$Pb | $^{190}$Hg | $^{194}$Hg |
|---|---|---|---|
| B(E1)$_{exp}$ | $1.0(2) \cdot 10^{-4}$ | $2 \cdot 10^{-3\,b)}$ | $10^{-5\,c)}$ |
| B(E1)$_{calc}^{a)}$ | $10^{-6}$ | $10^{-6} - 10^{-4}$ | $10^{-8} - 10^{-6}$ |

$^{a)}$ from [11, 12];   $^{b)}$ from [9];   $^{c)}$ from [13]

## 4. Summary

In SD $^{193}$Pb all the quasineutron Routhians that one can reasonably expect near the Fermi level have been observed. Calculated Routhians not identified in experiment may well lie at higher energies in more realistic mean-field calculations. Thus, $^{193}$Pb is the first nucleus in which a near-complete spectroscopy has been performed at superdeformation.

In an investigation of $^{196}$Pb the Ge Clover detectors of the EUROBALL $\gamma$-ray spectrometer array were used as Compton polarimeters to measure the linear polarization of transitions linking states of the second excited

SD band (band 3) to states of the yrast SD band. The identification of these transitions as enhanced electric dipoles gives strong evidence that the excited band is built on an octupole vibration. This measurement is an important step towards establishing that this phenomenon is wide spread in the A = 190 region of superdeformation.